\documentclass[letter]{aa} 
\usepackage{graphicx}
\usepackage{txfonts}
\begin{document}
\title{Periodic variability of 6.7\,GHz methanol masers in G22.357$+$0.066 
\thanks{Figures 3 and 4 are only available in electronic form at http://www.aanda.org} 
}

\author{M. Szymczak  \inst{1},
        P. Wolak \inst{1}, 
        A. Bartkiewicz \inst{1},
	\and H.J. van Langevelde \inst{2, 3}
	}
\institute{
	  Toru\'n Centre for Astronomy, Nicolaus Copernicus 
          University, Gagarina 11, 87-100 Toru\'n, Poland 
\and      Joint Institute for VLBI in Europe, Postbus 2, 7990 AA
          Dwingeloo, The Netherlands 
\and      Sterrewacht Leiden, Postbus 9513, 2300 RA Leiden, The Netherlands 
}


\date{Received 28 April 2011 / Accepted 19 May 2011 }

\abstract
{}
{We report the discovery of periodic flares of 6.7\,GHz methanol maser in the young massive 
stellar object G22.357+0.066.}
{The target was monitored in the methanol maser line over 20 months with the Torun 32\,m telescope. 
The emission was also mapped at two epochs using the EVN.}
{The 6.7\,GHz methanol maser shows periodic variations with a period of 179\,days. The periodic behavior
is stable for the last three densely sampled cycles and has even been stable over $\sim$12\,years, 
as the archival data suggest.
The maser structure mapped with the EVN remains unchanged at two epochs just at the putative flare maxima 
separated by two years. The time delays of up to $\sim$16\,days seen between maser features 
are combined with the map of spots to construct the 3-dimensional structure of the maser region. 
The emission originating in a single $\sim$100\,AU layer can be modulated by periodic changes in 
the infrared pumping radiation or in the free-free background emission from an HII region.}     
{}

\titlerunning{Periodic variability of the 6.7\,GHz methanol maser in G22.356$+$0.066}
 
\authorrunning{M. Szymczak et al.}

\maketitle

\section{Introduction}
Variability is a definite characteristic of methanol masers associated with massive
young stellar objects (MYSO). The observations of 6.7\,GHz maser flux in
several sources have shown a significant level of variability on timescales of 
a few days to several years (Caswell et al.\,\cite{caswell95}; MacLeod \& Gaylard\,\cite{macleod96};
Goedhart et al.\,\cite{goedhart04, goedhart09}; Sugiyama et al.\,\cite{sugiyama08}).
The possible causes of variability are changes in the pump rate or the maser path 
length, such as those caused by large scale motions (Caswell et al.\,\cite{caswell95}).
A long-term monitoring of 54 sources has revealed a diversity of types of behavior at 6.7\,GHz,
including periodic flares with periods of 132$-$520 days found in six objects (Goedhart et al.\,\cite{goedhart04}). 
For the best-studied periodic source G9.62$+$0.20E, van der Walt et al. (\cite{vanderwalt09})
propose a colliding wind binary that modulates the background radio continuum and/or 
pumping radiation with a period of 244 days. Araya et al. (\cite{araya10}) report the detection
of $\sim$237\,day periodicity in the 4.8\,GHz formaldehyde and 6.7\,GHz methanol maser lines 
towards MYSO candidate G37.55$+$0.20. As an alternative scenario for those variations,
they propose regular changes in the infrared pumping radiation due to periodic accretion 
onto a young binary system.

In this Letter we report the discovery of periodic variations in the 6.7\,GHz maser flux in G22.357$+$0.066.
The position of 6.7\,GHz maser (Schutte et al.\,\cite{schutte93}) measured with VLBI
(Bartkiewicz et al.\,\cite{bartkiewicz09}) coincides within 0\farcs05 with the water maser
(Bartkiewicz et al.\,\cite{bartkiewicz11}).
The source of 1.2\,mm dust continuum (Beuther et al.\,\cite{beuther02}) is offset by 2\farcs9
from the methanol maser. The extended ($\sim$23$\times$10\arcsec) HII region (White et al.\,\cite{white05})
and H110$\alpha$ recombination line and 4.8\,GHz formaldehyde absorption (Sewilo et al.\,\cite{sewilo04}) 
are offset by more than 10\arcsec, corresponding to 0.2\,pc for the near kinematic distance 
of 4.86\,kpc (Reid et al.\,\cite{reid09}).
Using the Torun 32\,m dish data we found periodic maser flares with a high degree of stability 
and were able to determine time delays between the individual spectral features. 
Those delays and high angular resolution maps obtained with the EVN are used to recover 
the three-dimensional structure of the maser emission.
There is convincing evidence for radiative excitation of maser spots by a variable central object.

\section{Observations}
The 6668.519\,MHz spectra were obtained with the Torun 32\,m radio telescope from 2009 June to 
2011 February at irregular intervals of one to six weeks. From 2010 September, when it became clear that 
the source was varying on shorter timescales, the monitoring interval was shortened to 1$-$20 days.
The half power beam width of the antenna was 5\farcm5, and the system temperature was typically 
around 40\,K. A frequency-switching mode was used. The signal of dual circular polarization was 
analyzed in a 4\,MHz bandwidth divided into 4096 channels, yielding a velocity resolution of 
0.05\,km\,s$^{-1}$ after Hanning smoothing. A typical rms noise level in the spectra after averaging 
the two polarizations was 0.2$-$0.3\,Jy. The flux density scale was established by observations of 
3C123 assuming flux density from Ott et al. (\cite{ott94}), while the stability of the system was 
regularly checked by observations of the non-variable source G32.745$-$0.076 
(Caswell et al.\,\cite{caswell95}). Careful inspection has shown that some features in this source 
are essentially constant within 2$-$3\% over the two years of monitoring. 
The accuracy of the absolute flux calibration was usually better than 10\%.  

Archival spectra were retrieved from the Torun 32\,m archive for MJD 1392, 2146, 4657, and 4849,
where MJD=JD$-$2450000. The spectral resolution was 0.05\,km\,s$^{-1}$, while the resulting accuracy
of absolute flux was about 15\%. For the first two epochs, the velocity scale was subject to 
$\pm$0.4\,km\,s$^{-1}$ error, and we assumed that the peak velocity was 80.1\,km\,s$^{-1}$.
The target was observed with the EVN at two epochs: 2007 June 13 and 2009 May 29. Details of the
observational setup, data reduction, and results for the first epoch are described in 
Bartkiewicz et al. (\cite{bartkiewicz09}). Uniformly weighted maps were obtained at both epochs
with the respective beam 8.7$\times$4.8\,mas and 6.0$\times$4.4\,mas and spectral resolution
of 0.18 and 0.09\,km\,s$^{-1}$. 

\section{Results}
\begin{figure}
   \resizebox{\hsize}{!}{\includegraphics[angle=0]{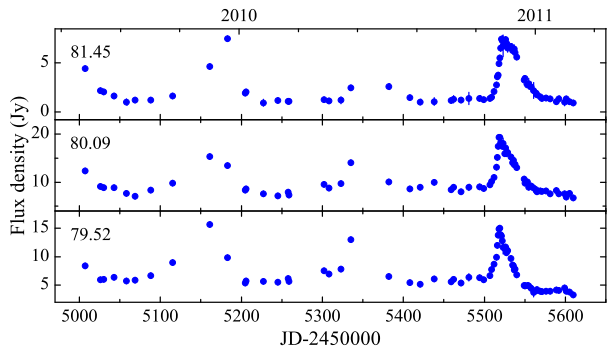}}
   \caption{Light curves of selected 6.7\,GHz methanol maser features in G22.357$+$0.066. 
    The 5 channel averages, labeled by the LSR velocity of the central channel, are shown.}
   \label{light-curves}
\end{figure}

The light curves of the three strongest features in the spectrum are displayed in 
Fig. \ref{light-curves}. Most of the spectral features show cyclic flares usually
followed by the return of the flux density to basically the same quiescent level 
(within $\sim$0.2\,Jy) between flares. The feature at 79.5\,km\,s$^{-1}$ exhibited 
a linear decrease (1.29\,Jy\,yr$^{-1}$) of the quiescent flux during the monitoring period.

\begin{figure}
   \resizebox{\hsize}{!}{\includegraphics[angle=-90]{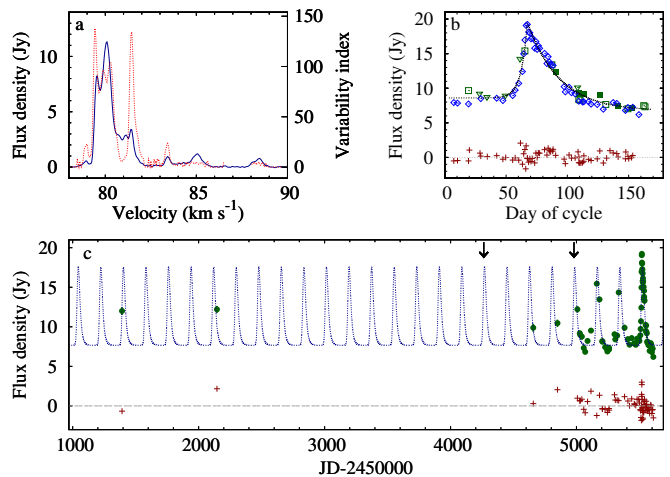}}
   \caption{{\bf a)} Average 6.7\,GHz maser line profile of G22.357+0.066 (blue line) with the variability 
    index $VI_1$ superimposed (red dotted line). {\bf b)} Folded flare profile at 80.1\,km\,s$^{-1}$ feature.
    The dotted line represents the best fit to the curve. The symbols mark different cycles. 
    {\bf c)} Time series for the 80.1\,km\,s$^{-1}$ feature with 
    the best-fit curve superimposed. The arrows mark the dates of the EVN observations.  
    The residuals from the fittings in {\bf b)} and {\bf c)} are shown by the crosses.}
   \label{flare}
\end{figure}

We quantified the variability of maser flux using the variability indices $VI_1$ and $VI_2$.
The first index is defined (Goedhart et al.\,\cite{goedhart04}) as

\begin{equation}
 VI_1 = {\big(}\sum_{i=1}^N(m(t_i) - \overline{m})^2 - \sum_{i=1}^N(n(t_i) - \overline{n})^2{\big)}{\big/}\overline{m} 
\end{equation}

\noindent
where $N$ is the total number of observations, $m(t_i)$ the observed flux density
at a given spectral channel at $t_i$ epoch, $n(t_i)$ the flux density at any 
emission-free spectral channel at $t_i$ epoch, and
$\overline{m}$ and $\overline{n}$ are the average flux densities. 
The second variability index is  $VI_2=(S_{\rm max} -S_{\rm min})/(S_{\rm max} +S_{\rm min})$, 
where $S_{\rm max}$ and $S_{\rm min}$ are the highest and lowest flux densities, respectively. 
The spectrum of $VI_1$ is superimposed on the 6.7\,GHz flux density spectrum (Fig. \ref{flare}a). 
The values of $VI_1$ and $VI_2$ for the most recognizable features are given in Table \ref{table1}.
We notice that the blue-shifted emission, i.e. at velocities lower than the systemic velocity of 
84.2\,km\,s$^{-1}$ (Szymczak et al.\,\cite{szymczak07}), is significantly more variable than 
the red-shifted emission.

\begin{table}
\caption {Variability parameters of the maser features in G22.357$+$0.066.}  
\label{table1}
\begin{tabular}{c r c c c}
\hline
\hline
Velocity & $VI_1$ & $VI_2$ & FWHM(SE) & $\Delta$t \\
(km\,s$^{-1}$) &   &       & (day) & (day) \\   
\hline
78.95 &   16.1 & 0.87 & 14.5(1.2) & $-$0.9(0.7)  \\
79.52 &  104.0 & 0.48 & 13.9(0.9) & $-$3.4(0.2)  \\
80.09 &   85.1 & 0.39 & 19.6(0.7) & 0 (0.2)      \\
81.45 &  111.1 & 0.71 & 24.8(0.8) & 5.4(0.3)     \\
83.42 &   18.0 & 0.64 & 23.5(1.6) & 3.9(0.6)     \\
85.05 &    1.4 & 0.13 &  -        &{\it 4.8(1.7)}     \\
88.47 &    2.0 & 0.30 &  -        &{\it 13.2(3.0)}    \\
\hline\end{tabular} 
\end{table} 

Figure \ref{flare}c shows the light curve of the 80.1\,km\,s$^{-1}$ feature. The new data are supplemented
by the archival 32m telescope observations at four epochs (see Sect.2). No velocity shift of the strongest 
feature (80.1\,km\,s$^{-1}$) was detected within the velocity resolution for ours 
and the Walsh et al. (\cite{walsh97}) data. The flux variations are fitted using the
equation (David et al.\,\cite{david96}): 
$S(t)=A^{s(t)} + C$, where $A$ and $C$ are constants and 
$s(t)= [(b $cos$(\omega t + \phi))/(1 - f $sin$(\omega t +\phi)] + a$.
Here, $b$ is the amplitude measured relative to the average value $a$, $\omega=2\pi/P$, $P$ is the period,
$\phi$ the phase, and $f$ the asymmetry parameter defined as the rise time from the minimum to the 
maximum flux, divided by the period. The best fit to the observational points is shown in Fig.\ref{flare}c,
and the inferred period is 179.2$\pm$0.6\,days. The time of flux maximum of the last cycle was MJD=5518.8$\pm$1.6.
Small residuals ($<$2.2\,Jy) for the 1999 and 2000 points  
suggest the high regularity of flares on a timescale of $\sim$12\,years. Furthermore, a peak flux of 13\,Jy
reported for 1994 April 8-12 observation (Walsh et al.\,\cite{walsh97}), i.e., 16-20\,days before 
a back-extrapolated maximum supports this suggestion. 

Figure \ref{flare}b shows the shape of the flare at feature 80.1\,km\,s$^{-1}$. The data are folded on
the precise 179.2-day period, and the origin of time axis is arbitrary. 
The rise and decay times of the flare were 14.5$\pm$1.7 and 42.9$\pm$1.7 days, respectively.  
The full width of flare at half maximum (FWHM) was 19.6$\pm$0.7\,days. The values of FWHM for other features are given
in Table \ref{table1}. There is a slight increase in the flare duration for features with higher radial velocities.
This tendency is also visible in the data from the last cycle (Fig. \ref{dynamic-spectra}); 
for the blue-shifted emission; lower than 80.5\,km\,s$^{-1}$, the flare lasted on average 65.8$\pm$2.6\,days
and increased up to $\sim$80\,days for 81.01 and 81.45\,km\,s$^{-1}$ features. 

\onlfig{3}{
\begin{figure*}
   \resizebox{\hsize}{!}{\includegraphics[angle=-90]{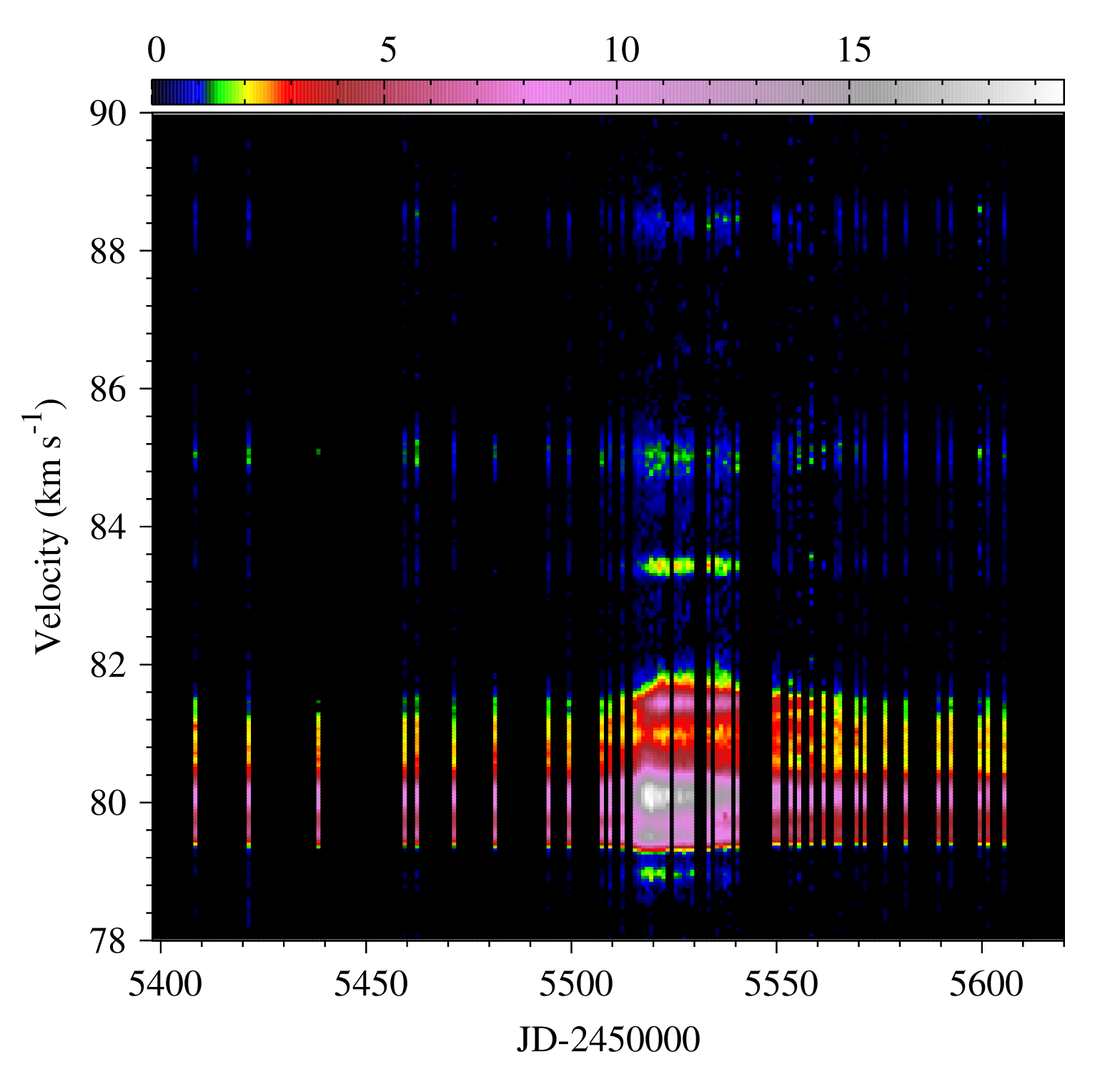}}
   \caption{6.7\,GHz flux density of G22.357$+$0.066 as a function of time and velocity.
    The timescale covers the last cycle of variability from September 2010 to February 2011.}  
   \label{dynamic-spectra}
\end{figure*}
}

The most striking characteristic of the flare are the time delays of the maximum flux between features of 
different velocities (Fig. \ref{dynamic-spectra}). 
To measure the time delay precisely, we used the discrete autocorrelation function 
(Edelson \& Krolik\,\cite{edelson88}), because the light curve is covered with unevenly sampled
data. The calculated time delays, $\Delta$t, for the central velocities of maser components are listed in 
Table \ref{table1}. A negative lag means that the maser feature flares before the reference feature 
(80.1\,km\,s$^{-1}$).
The delays range from -3.4 to 13.3\,days. The estimated uncertainties in the time delays are usually 
below 0.7\,days but are higher than 1.5\,days for the components, marked in italics, that have a low 
signal-to-noise ratio and small flare amplitudes.

The 6.7\,GHz maser structure of the target observed with the EVN at MJD 4981, i.e. exactly at the 
time of the putative maximum of the flare (Fig. \ref{flare}b), is shown in Fig. \ref{evn-maps}. 
For comparison, we present the map obtained by Bartkiewicz et al. (\cite{bartkiewicz09}). 
Their observations were also carried out just at a flare maximum (MJD 4264), as estimated from
the 32m dish data. In general the structure of the maser emission is well preserved over two years, as 
the relative positions of the spots are the same within $\sim$0.5\,mas. There are only two spots for
which the morphology has apparently changed. The cluster of spots centered at 84.86\,km\,s$^{-1}$ was more 
scattered at the first epoch. No emission from the S-E part of this cluster was detected at
the second epoch. Similarly, the emission from the northernmost part of the cluster of spots 
near 88.47\,km\,s$^{-1}$ disappeared after two years. These two red-shifted clusters show relatively 
weak emission ($<$1.7\,Jy). Their flare profiles are flatter and broader than those of 
the blue-shifted emission (Fig. \ref{dynamic-spectra}). One can suppose that they are diffuse 
and have low brightness, which can be filtered out more at the second epoch when observed with 
a beam 1.6 times smaller and with spectral resolution 2 times higher than at the first epoch. 
No new maser spots are formed over two years. The brightness of individual maser spots of the blue-shifted
emission usually changes by less than 1$-$2\,Jy, which is within the error range of absolute amplitude
calibration. We conclude that there are no significant differences in the source structure at the two epochs.

\onlfig{4}{
\begin{figure*}
   \resizebox{5.5in}{!}{\includegraphics[angle=0]{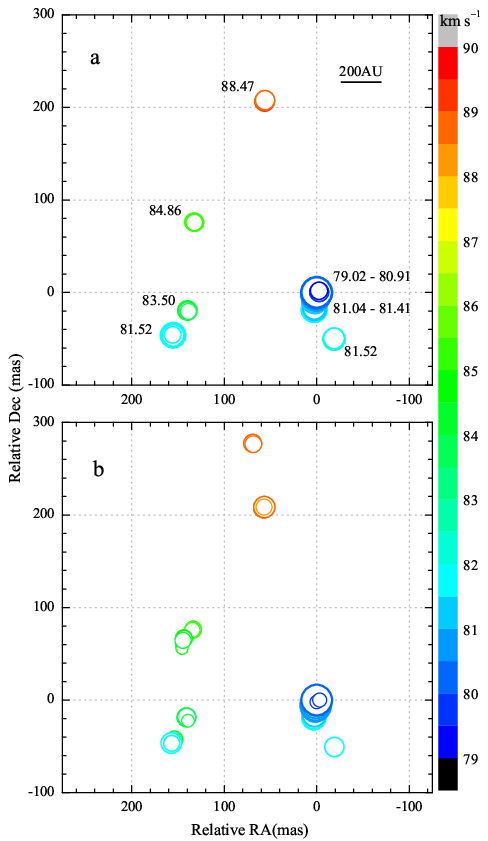}}
   \caption{Maps of 6.7\,GHz methanol maser in G22.357+0.066 obtained with the EVN at two epochs; 
    MJD 4981 ({\bf a}) and MJD 4264 ({\bf b}, adapted from Bartkiewicz et al.\,\cite{bartkiewicz09}).
    The coordinates are relative to the brightest spot. The sizes of symbols are proportional to the logarithm
    of the brightness of maser spots. The colors of circles relate to the LSR velocity scale presented in the 
    wedge. The groups of spots are also labeled by their central velocities or the velocity ranges. The linear
    scale, derived for the distance of 4.86\,kpc, is shown by the horizontal bar.}
   \label{evn-maps}
\end{figure*}
}

\section{Discussion}
The detection of periodic variations in the 6.7\,GHz methanol maser flux in G22.357+0.066 extends a class
of periodic maser sources (Goedhart et al.\,\cite{goedhart04}). We note that the flare profile of the target 
is very similar to what is reported in G9.62+0.20E at 12.2\,GHz (van der Walt et al.\,\cite{vanderwalt09}).
Specifically, the ratio of the rise to the decay time is 0.34 in G22.357+0.066 and 0.37 in G9.62+0.20E,
and the respective values of the relative amplitude of flux density variation are 1.5 and 2.0.
This suggests that the same underlying processes drive the periodicity in the two sources. 

The cause of the periodic flares of methanol maser sources is still unknown. Several possibilities have
been considered (Goedhart et al.\,\cite{goedhart09}; van der Walt et al.\,\cite{vanderwalt09}; 
Araya et al.\,\cite{araya10}; van der Walt \cite{vanderwalt11}) including (i) disturbances of the masing region 
by shock waves or clumps, (ii) variations in the background radio continuum and/or of pump photons due to stellar 
pulsations, modulations of the radiation from a colliding wind binary, or periodic accretion of circumbinary disk
material. For G22.357+0.066 we found that all the features return their 6.7\,GHz flux to basically the
same quiescent level between flares and that the velocities of different features and the spectrum shape are stable 
over at least 12\,years. Moreover, the spatial structure on a mas scale remains essentially the same for the two
flares spanning two years. These facts imply that the velocity and spatial structure of the source remain unaffected 
by whatever mechanism underlies the periodic variation. Furthermore, the maximum projected separation 
of 1050\,AU between the maser spots in the source, for the distance of 4.86\,kpc, 
is well within a range of typical distances between the maser region and the driving MYSO
(Bartkiewicz et al.\,\cite{bartkiewicz09}). If the flares are triggered by a propagating density wave or shock front,
its velocity should be higher than 10$^4$\,km\,s$^{-1}$ to account for 179 day periodicity. Such a fast shock would
easily dissociate the molecules (Hollenbach \& McKee \,\cite{hollenbach89}). Therefore we exclude shock waves 
as a possible mechanism of the flaring.

The time delays between flares of individual features, with the highest values of $\sim$16 days between the flare 
in the blue-shifted feature at 79.5\,km\,s$^{-1}$ and the extreme red-shifted feature at 88.5\,km\,s$^{-1}$, clearly 
implies the radiative coupling of a single triggering process with the maser regions.
This is consistent with very similar flare profiles for the emission features which arise from single 
cluster of spots, i.e. at velocity lower than $\sim$81\,km\,s$^{-1}$. 
However, the flare profile of the 81.5\,km\,s$^{-1}$ feature is different 
(Figs. \ref{light-curves}, \ref{dynamic-spectra}). This feature is a blend of emission from the two spatially 
separated clusters (Fig. \ref{evn-maps}), and its blurred flare profile is likely due to the different time delays 
of each cluster. 
The red-shifted features near 85.0 and 88.5\,km\,s$^{-1}$ have flat and poorly pronounced 
flare profiles. Assuming that the maser flaring is caused by changes in the infrared pumping radiation,
the red-shifted emission may come from the far side of the envelope and it is less bright and less saturated 
due to the lack of background radiation. Alternatively, it can be due to either strongly directional 
amplification of the maser emission, where backward propagation dominates, or blending of low 
intensity extended emission.  However, if the flaring of all the maser features is due to changes in 
the free-free emission of a background HII region, then all the maser features must be on the near side of
the HII region and no simple geometry of the maser region can be recovered from the time delay measurements.    

\begin{figure}
   \resizebox{3.70in}{!}{\includegraphics[angle=-90]{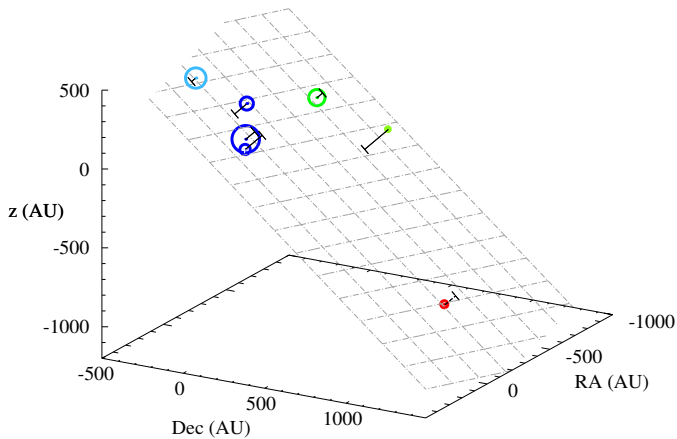}}
   \caption{Three-dimensional view of 6.7\,GHz maser in G22.357$+$0.066. The z axis is towards the observer.
    The symbol sizes are proportional to the variability index $VI_2$ (Table \ref{table1}),
    and the colors mark the LSR velocity scale as in Fig. \ref{evn-maps}. The bars mark the distance of spots 
    to the plane shown by the grid.}
   \label{structure}
\end{figure}

We used the time delays and the EVN map to construct the three-dimensional structure of the maser region.
For a scenario of modulation of the infrared pumping, we assumed that a source triggering the periodic variations 
coincides with the line of sight to the strongest feature at 80.1\,km\,s$^{-1}$ and that the maser emission arises 
in a {\it spherical} envelope of mean radius $r_0$. The value of $r_0$ is calculated by minimizing residuals of 
the least-square fits of the envelope to the positions of spots. 
The best fitted value of $r_0$ is 710$\pm$140\,AU, and the maser spots lie in a layer of
$\sim$100\,AU thickness inclined 47\degr\, from the line of sight (Fig. \ref{structure}). The overall size of
the maser structure is $\sim$1500$\times$600\,AU. It is a somewhat surprising result, which may indicate that
the maser emission forms in a circumstellar disc or torus. We note that this is not unique model because
neither the location of the triggering source nor details on the pump rate variations are known.

The strongest maser spot coincides within 0\farcs62 with the GLIMPSE source G022.3564+00.0662 
(Fazio et al.\,\cite{fazio04}); i.e., the total emission seen by the EVN falls within one {\it Spitzer} 
pixel of nominal size $\sim$1\farcs2.
The colors of this counterpart source are typical of an MYSO (Cyganowski et al.\,\cite{cyganowski09}).
No compact 5\,GHz radio continuum ($1\sigma$=0.34\,mJy\,b$^{-1}$) or any H$_2$CO absorption was detected 
with the beam $1\farcs6\times1\farcs1$ (Araya 2010, priv. comm.). The extended ($\sim$23\arcsec$\times$10\arcsec) 
HII region (White et al.\,\cite{white05}) and H110$\alpha$ recombination line and 4.8\,GHz formaldehyde absorption 
(Sewilo et al.\,\cite{sewilo04}) are offset by more than 10\arcsec, while a faint (1.02\,mJy\,beam$^{-1}$) 
and extended 8.4\,GHz emission is located by 14\farcs8 to the N$-$E of the source (van der Walt\,\cite{vanderwalt03}), 
corresponding to 0.2$-$0.3\,pc. There is a suggestion that the maser lies at the edge of an evolved HII region of 
$\sim$5\arcmin\, size that may trigger a new generation of stars (van der Walt et al.\,\cite{vanderwalt03}). 
Therefore it is probable that more sensitive observations will identify a source triggering periodic variation 
of the methanol maser.

A model of a colliding wind binary was proposed to explain the best-studied periodic source G9.62+0.20E
(van der Walt et al.\,\cite{vanderwalt09}; van der Walt \cite{vanderwalt11}). 
It provides the mechanism of periodic changes in the background
radio continuum and infrared radiation. Regularly modulated pulses of ionizing radiation passing through 
a volume of partially ionized gas against which the maser is projected produce periodic changes in the electron 
density so that the background free-free emission is variable. 
Within the frame of that model, the observed decay time of the 12.2\,GHz maser flare in G9.62+0.20E 
is consistent with the characteristic recombination time of a hydrogen plasma with densities of 
$10^5-10^6$\,cm$^{-3}$. As there is a hint of an extended radio continuum source in the field in the case 
of G22.357+0.066, it is reasonable to apply the van der Walt at al. model. 
We fitted eq. 3 (van der Walt et al.\,\cite{vanderwalt09}) to the decay part of the 6.7\,GHz flare
in the source (see Fig. 2b). The best fit was obtained for the equilibrium electron density ranging from
5$\times$10$^4$ to 1$\times$10$^5$\,cm$^{-3}$ and the electron density at the decay start time of 
1.03$\times$10$^6$\,cm$^{-3}$. These values are well within the ranges reported for G9.62+0.20E 
(van der Walt et al.\,\cite{vanderwalt09}) where two components of the decay of the flare profile were
fitted. This suggests that the maser flare in G22.357+0.066 can be due to changes in the number of free-free 
background photons from an HII region.

Araya et al. (\cite{araya10}) propose that periodic heating of the dust and increasing the infrared radiation 
can be caused by accretion of material from the circumbinary disk onto the protostars or accretion disks.
As pointed out in van der Walt (\cite{vanderwalt11}), this scenario is less probable for periodic masers, because 
for a system with an eccentricity of 0.1, a significant time lag is expected in the accretion rate onto 
the two stars, and variations in the total accretion rate will not show the sinusoidal-like pattern. 
For a highly eccentric system, the accretion rate changes rapidly near periastron 
(Artymowicz \& Lubow\,\cite{artymowicz96}).
Although the scenarios of the colliding wind binary and periodic accretion onto a young binary system provide 
a periodic source of photons heating the circumstellar dust, thereby affecting the pumping infrared
radiation field, the typical cooling time for the optically thick case of a few days 
(van der Walt et al.\,\cite{vanderwalt09}) is too short to explain the decaying part of the maser flare profile.         
Monitoring of the source in the radio and infrared will provide major progress in explaining the causes of
its periodic variability. 

\begin{acknowledgements}
We thank the anonymous referee for useful comments that have improved the paper.
The work was supported by the Polish Ministry of Science and Higher Education through grant N N203 386937.
The European VLBI Network (EVN) is a joint facility of European, Chinese, South African, and other 
radio astronomy institutes funded by their national research councils
\end{acknowledgements}

\end{document}